\documentstyle[12pt,epsfig,amsmath,amssymb,pstricks]{article}
\begin{document}

\title{\bf Bell's Theorem from Moore's Theorem}

\author{{Chris Fields}\\ \\
{\it 815 East Palace \# 14}\\
{\it Santa Fe, NM 87501 USA}\\ \\
{fieldsres@gmail.com}}
\maketitle

\begin{abstract}
It is shown that the restrictions of what can be inferred from classically-recorded observational outcomes that are imposed by the no-cloning theorem, the Kochen-Specker theorem and Bell's theorem also follow from restrictions on inferences from observations formulated within classical automata theory.  Similarities between the assumptions underlying classical automata theory and those underlying universally-unitary quantum theory are discussed.
\end{abstract}



\section{Introduction}

In 1956, Edward F. Moore proved that no finite sequence of observations of the input-output behavior of a classical finite state machine (FSM) is sufficient to uniquely identify the FSM \cite{moore:56}; similar results were published around the same time by Ashby \cite{ashby:56} and others.  Moore, whose primary interest was in reverse engineering, remarked that this result ``means that it will never be possible to perform  experiments on a completely unknown machine which will suffice to identify it from among the class of all sequential machines'' (p. 140) but did not elaborate on the implications of this for experimental physics.  Any finite sequence of explicitly-specified discrete states of any system can, however, be described formally as an execution trace of a classical FSM \cite{hopcroft:79}; in such a description, the states of the FSM are identified with the observed system states, and the transition rule of the FSM is specified explicitly as an ordered list of the observed state-to-state transitions.  Any finite sequence of observations of discrete states of any physical system $\mathbf{S}$ for which the observational outcomes are explicitly recorded as classical information can, therefore, be described formally as a sequence of observations of a classical FSM implemented by $\mathbf{S}$.  Moore's theorem shows that no such finite sequence of classically-recorded experimental observations of discrete states of $\mathbf{S}$ is sufficient to uniquely specify \textit{which} classical FSM $\mathbf{S}$ implements, as an arbitrarily-large number of classical FSMs are consistent with any given finite sample of state transitions.  Moore's theorem is, therefore, a classical forerunner of later ``no-go'' results that characterize observations of the physical world: it demonstrates an experimentally-unresolvable ambiguity in the formal characterization of any physical system as a classical FSM on the basis of a finite number of classically recorded observational outcomes.

This paper examines Moore's theorem in the context of quantum theory.  It shows that Moore's theorem together with the assumption that all information is at all times physically encoded \cite{landauer:99} blocks two assumptions that are commonly made regarding finite sequences of classically-recorded observational outcomes: the assumption that the system that has been observed can be fully characterized by a specified collection of physical degrees of freedom and hence a specified Hilbert space, and the assumption that the recorded observational outcomes all characterize a single system.  These assumptions are blocked by Moore's theorem regardless of whether the ``system'' referred to is macroscopic or microscopic, observed directly or observed through the use of an apparatus.  The paper then shows that the inferences from observational outcomes that are blocked by the well-known ``no-go'' theorems of Wooters and Zurek \cite{wooters:82}, Kochen and Specker \cite{kochen:67} and Bell \cite{bell:64} are instances of inferences that are blocked by Moore's theorem.  These no-go results can, therefore, be regarded as establishing in particular cases the restrictions on the inferences that can be drawn from finite sequences of classically-recorded observational outcomes that are established more generally by classical automata theory.  That this should be so is \textit{prima facie} surprising: Moore's theorem follows from an analysis of classical observation, while the quantum no-go theorems follow from the assumption of unitary dynamics.  Their correspondence suggests that the inability to place upper bounds on the degrees of freedom of a system, and hence the inability to rule out dynamical effects of previously-unobserved degrees of freedom that is entailed by Moore's theorem has the same consequences for observation as does the existence of quantum entanglement.

\section{Moore's theorem in the context of quantum theory}  

Moore's second theorem states:

\begin{quote}
``Given any machine $\mathbf{S}$ and any multiple experiment performed on $\mathbf{S}$, there exist other machines experimentally distinguishable from $\mathbf{S}$ for which the original experiment would have had the same outcome.''
\begin{flushright}
(\cite{moore:56} p. 140)
\end{flushright} 
\end{quote}

For Moore, a ``machine'' is a classical, deterministic FSM: a finite set of explicitly-specified states together with a deterministic transition rule that specifies, either explicitly or implicitly, all of the state transitions that can be executed by the machine.  To experimentally characterize an FSM is to determine, by sequentially placing the machine in observationally-distinct states and recording the state transitions that follow each such manipulation, the complete transition rule.  Moore's theorem concerns an unknown machine - the object of any exercise in reverse engineering - that is assumed to be a deterministic FSM but for which neither the list of possible states nor the transition rule is available.  Moore proves his theorem by noting that any transition rule that is inferred from a finite number of observations, and which therefore specifies sequences of state transitions starting from a finite number of observationally-distinct initial states, could either be the \textit{complete} transition rule of an FSM defined over \textit{only} the states explicitly characterized by the observations, or a \textit{partial} transition rule of a larger FSM defined over a larger number of states.  Hence Moore's theorem shows that even in the case of classical FSMs, no finite set of observations is sufficient to conclusively identify the state space, and hence the behavioral degrees of freedom, of the system being observed.  In particular, Moore's theorem shows that while finite observations can establish that a system has particular degrees of freedom, and hence can put a lower bound on the size of the system's state space, they cannot establish that a system has \textit{only} those degrees of freedom, and hence cannot put an upper bound on the size of the system's state space.

It is important to emphasize that Moore's theorem in no way blocks either the construction of theoretical models of the behavior of a system that assume that the system has the degrees of freedom that experimental observations indicate, or the design of further experimental manipulations based on such models.  Hence Moore's theorem does not challenge the \textit{de facto} methods of experimental physics, or of any other science.  What Moore's theorem does challenge is the theoretical assumption that a system has \textit{only} the degrees of freedom that have been characterized by experimental observations: it challenges the assumption that any theoretical model is complete.  Such completeness assumptions are typically implicit, as they are typically embedded in and hence unavoidable consequences of employing the formalism with which theoretical models are constructed.  The primary practical consequences of Moore's theorem are, therefore, restrictions on the assumptions that can be built into a model-building formalism.  It will be shown in \S 4 below that these restrictions encompass those imposed by the quantum no-go theorems; the present section examines these restrictions and shows how imposing them alters the representation of observations within the standard quantum-mechanical formalism.  

As Moore's theorem concerns what can be inferred from observations, it is useful to examine it using an explicit physical model of observation.  Following Landauer \cite{landauer:99}, let us explicitly assume that all information is physically encoded.  Suppose an observer $\mathbf{O}$ is embedded in an environment $\mathbf{E}$, that $\mathbf{O}$ extracts classical information from $\mathbf{E}$ in the form of $N$ discrete observational outcomes that are obtained at $N$ distinct times $t_{1} ... t_{N}$, and that these outcomes are recorded by $\mathbf{O}$ as a finite sequence of classical symbols $k_{1} ... k_{N}$.  These $N$ classical records can be regarded as completely specifying $N$ states of a classical FSM, and the $N-1$ transitions $k_{j} \rightarrow k_{j+1}$ can be regarded as a complete, explicit specification of the transition rule of this FSM.  Moore's theorem then applies, and states that the recorded symbols $k_{1} ... k_{N}$ and transitions $k_{j} \rightarrow k_{j+1}$ could instead be only partial specifications of any of an arbitrarily large number of distinct classical FSMs.  Moore's theorem blocks, in particular, the ``Occam's razor'' inferences that the symbols $k_{1} ... k_{N}$ constitute a complete specification of the state space of $\mathbf{E}$ and that the transitions $k_{j} \rightarrow k_{j+1}$ constitute a complete specification of the dynamical behavior of $\mathbf{E}$, regardless of the size of $N$.

Nothing has been specified, in the above, about the interaction between $\mathbf{O}$ and $\mathbf{E}$ except that it results in the encoding by $\mathbf{O}$ of the classical records $k_{1} ... k_{N}$.  Let us suppose that $\mathbf{O}$ and $\mathbf{E}$ are described by distinct collections of quantum degrees of freedom $\lbrace o_{i} \rbrace$ and $\lbrace e_{j} \rbrace$ respectively, and hence that their interaction can be represented by a Hamiltonian $H_{\mathbf{OE}} = \mathit{\sum_{ij}  H_{ij}}$, where $H_{ij}$ couples the $i^{th}$ degree of freedom of $\mathbf{O}$ to the $j^{th}$ degree of freedom of $\mathbf{E}$.  As Moore's theorem blocks the inference that the symbols $k_{1} ... k_{N}$ constitute a complete specification of the state space of $\mathbf{E}$, it clearly blocks the inference that the symbols $k_{1} ... k_{N}$ provide sufficient information to completely specify $H_{\mathbf{OE}}$, again regardless of the size of $N$.

It is standard, in quantum theory, to represent interactions between observers and their environments not by Hamiltonians, but rather by generalized observables, i.e. positive operator-valued measures (POVMs; \cite{nielsen-chaung:00} Ch. 2).  Let us suppose, therefore, that the $N$ classical records $k_{1} ... k_{N}$ encode $N$ outcomes $E_{k_{1}}^{\mathbf{E}} ... E_{k_{N}}^{\mathbf{E}}$ obtained by deploying a POVM $\lbrace E_{i}^{\mathbf{E}} \rbrace$ defined over the Hilbert space $\mathcal{H}_{\mathbf{E}}$ of $\mathbf{E}$.  As Moore's theorem blocks the inference that the symbols $k_{1} ... k_{N}$ constitute a complete specification of $\mathcal{H}_{\mathbf{E}}$, it clearly blocks the inference that the symbols $k_{1} ... k_{N}$ provide sufficient information to completely specify the deployed POVM $\lbrace E_{i}^{\mathbf{E}} \rbrace$.

The above restrictions on inferences from a sequence of classically-recorded observational outcomes $k_{1} ... k_{N}$ make no particular assumptions about the size or structure of $\mathbf{E}$; indeed $\mathbf{O}$'s inferences from $k_{1} ... k_{N}$ would be restricted by Moore's theorem even if $k_{1} ... k_{N}$ as a matter of fact \textit{was} a complete specification of the state space of $\mathbf{E}$.  The restrictions on inferences from a sequence of classical outcomes $k_{1} ... k_{N}$ that are imposed by Moore's theorem therefore apply not only to $\mathbf{O}$'s interactions with $\mathbf{E}$ as a whole, but also to $\mathbf{O}$'s interactions with any components of $\mathbf{E}$, i.e. to $\mathbf{O}$'s interactions with any subsets of the degrees of freedom $\lbrace e_{j} \rbrace$.  If $\mathbf{S}$ comprises such a subset of the quantum degrees of freedom of $\mathbf{E}$, Moore's theorem blocks any inference that a set of classical symbols $k_{1} ... k_{N}$ recorded in the course of interactions between $\mathbf{O}$ and $\mathbf{S}$ provides sufficient information to completely specify the Hilbert space $\mathcal{H}_{\mathbf{S}}$, the Hamiltonian $H_{\mathbf{OS}}$, or the POVM $\lbrace E_{i}^{\mathbf{S}} \rbrace$ that produced $k_{1} ... k_{N}$ as output.  

A case of particular interest is that in which $\mathbf{O}$ employs one subset of the degrees of freedom of $\mathbf{E}$, those describing an ``apparatus'' $\mathbf{A}$, to make measurements of another subset of the degrees of freedom of $\mathbf{E}$, those describing a ``system of interest'' $\mathbf{S}$.  In this case, $\mathbf{O}$ employs some POVM $\lbrace E_{i}^{\mathbf{A}} \rbrace$ to observe $\mathbf{A}$, and regards $\mathbf{A}$ as implementing a POVM $\lbrace E_{i}^{\mathbf{S}} \rbrace$ that acts on $\mathbf{S}$.  This latter POVM yields observational outcomes $k_{1} ... k_{N}$ that are classically recorded either by $\mathbf{A}$ directly or by $\mathbf{O}$ through the use of $\lbrace E_{i}^{\mathbf{A}} \rbrace$.  If $\mathbf{O}$ cannot completely specify $\mathcal{H}_{\mathbf{A}}$ by observing $\mathbf{A}$ with $\lbrace E_{i}^{\mathbf{A}} \rbrace$, $\mathbf{O}$ clearly cannot completely specify the Hamiltonian $H_{\mathbf{AS}}$ via which $\mathbf{A}$ interacts with $\mathbf{S}$, and so cannot completely specify the POVM $\lbrace E_{i}^{\mathbf{S}} \rbrace$ that $\mathbf{A}$ implements.  In particular, $\mathbf{O}$ cannot specify on the basis of a sequence of classically-recorded observational outcomes $k_{1} ... k_{N}$ generated by $\mathbf{A}$ that $H_{\mathbf{AS}} = H_{\mathbf{A \xi}}$ for some particular quantum degree(s) of freedom $\xi$ within $\mathbf{E}$, and hence cannot infer that the POVM $\lbrace E_{i}^{\mathbf{S}} \rbrace$ implemented by $\mathbf{A}$ is definable over $\mathcal{H}_{\xi}$.

It is commonplace in applications of quantum theory to \textit{stipulate} that some sequence of classically-recorded observational outcomes $k_{1} ... k_{N}$ was generated by the action of a specified POVM $\lbrace E_{i}^{\mathbf{S}} \rbrace$ defined over a specified Hilbert space $\mathcal{H}_{\mathbf{S}}$.  Such stipulations may in some cases reflect explicit hypotheses: it may be being hypothesized that $k_{1} ... k_{N}$ were generated by the action of $\lbrace E_{i}^{\mathbf{S}} \rbrace$, or that the observed system has some particular set of degrees of freedom and hence some particular $\mathcal{H}_{\mathbf{S}}$.  Moore's theorem clearly allows such hypotheses; they are essential to the practice of science, and are falsified if the involvement of additional degrees of freedom in the behavior of $\mathbf{S}$ is observationally confirmed.  What Moore's theorem blocks is treating it as a \textit{physical fact} that $k_{1} ... k_{N}$ were generated by the action of a particular $\lbrace E_{i}^{\mathbf{S}} \rbrace$ defined over a particular $\mathcal{H}_{\mathbf{S}}$ smaller than the environment in which the observer is embedded.  Zurek \cite{zurek:03}, for example, proposes as ``axiom(o)'' of quantum theory that ``the Universe consists of systems'' (p. 746) with which observers interact; ``quantum Darwinism'' \cite{zurek:06, zurek:09} is based on the assumptions that decoherence acts on these specific systems to create stable environmental encodings, and that it is these encodings that observers detect \cite{fields:10, fields:11}.  \textit{Defining} a POVM $\lbrace E_{i}^{\mathbf{S}} \rbrace$ over a specific Hilbert space $\mathcal{H}_{\mathbf{S}}$ embeds this assumption that observable systems exist as discrete entities in an observer-independent way into the quantum formalism.  Such a definition renders every observation system-specific: observing the position of a particular desk $\mathbf{D}$, for example, requires deploying a position observable $\hat{x}^{\mathbf{D}}$ defined over the Hilbert space of that desk, while observing the position of its associated chair $\mathbf{C}$ requires deploying an $\mathcal{H}_{\mathbf{C}}$-specific operator $\hat{x}^{\mathbf{C}}$.  Within this formalism, any sequence of outcomes $k_{1} ... k_{N}$ is guaranteed by definition to be generated by interactions between $\mathbf{O}$ and the particular degrees of freedom over which the deployed POVM is defined; the possibility that any other degrees of freedom are involved in any way in the production of the observed outcomes is ruled out by fiat.  The question with which Moore's theorem is concerned, that of how observers \textit{identify} a particular collection of degrees of freedom as a target of observations can no longer be raised: the identification of systems of interest as particular collections of physical degrees of freedom - i.e. as particular Hilbert spaces - is no longer a hypothesis, but is simply taken for granted whenever a specific POVM is employed to represent the process of observation.  This ``taking for granted'' represents an enormous step beyond the standard \cite{vonNeumann:32} first axiom of quantum theory: it is the step from ``this observable system has \textit{some} (to be determined) Hilbert space'' to ``this observable system has, by definition, \textit{this} Hilbert space.''

Moore's theorem shows that this common assumption that system identification can be taken for granted cannot be empirically justified, even in principle: no finite sequence of classically-recorded observations, no matter how large, is sufficient to completely specify the collection of physical degrees of freedom, and hence the ``system'' with which the observer has interacted over the course of the observations.  Only one formal response to this predicament is consistent with both the fundamental assumption that observers can obtain classical information only by deploying POVMs, i.e. only by undergoing physical interactions with their environments, and the formal requirement that POVMs be well-defined.  It is to treat any observer as equipped with a finite number of POVMs $\lbrace E_{i}^{\mathbf{X}} \rbrace$, $\lbrace E_{i}^{\mathbf{Y}} \rbrace$, $\lbrace E_{i}^{\mathbf{Z}} \rbrace$, etc., all of which are defined over the complete Hilbert space $\mathcal{H}_{\mathbf{E}}$ of the environment $\mathbf{E}$ in which the observer is embedded.  If $\mathbf{O}$ is equipped only with POVMs defined over \textit{all} of $\mathbf{E}$, $\mathbf{O}$ has no choice but to identify as a ``system'' $\mathbf{S}$, for example, \textit{whatever} collections of degrees of freedom of $\mathbf{E}$ produce recordable outcomes when acted on by some available POVM $\lbrace E_{i}^{\mathbf{S}} \rbrace$.  With this alternative formalism, all observations are treated as system-\textit{non}specific: observing the position of a desk involves deploying a generic position observable $\hat{x}^{\mathbf{desk}}$ and identifying as a ``desk'' whatever produces a recordable outcome.  If the POVMs deployed by an observer incorporate multiple commuting observables and hence yield as outcomes classical symbols that describe observations along multiple dimensions, their \textit{de facto} specificity in picking out systems of an intuitively coherent type may be high, but Moore's theorem blocks the inference that any finite POVM has unique specificity.  This formal approach of treating \textit{all} systems, whether microscopic or macroscopic, as identified solely by fixed sets of classically-recordable observational outcomes - i.e. as identified in the same way that electrons or quarks are identified - has been proposed elsewhere on the basis of a formal model of observers as finite information processing systems \cite{fields:12a}.  It is employed in what follows to show that the restrictions on inferences from observations that are imposed by the no-cloning theorem, the Kochen-Specker theorem and Bell's theorem are special cases of those imposed by Moore's theorem.

\section{Moore's theorem implies an observable-dependent exchange symmetry}

Suppose an observer $\mathbf{O}$ is equipped with a Geiger counter with a 2 cm diameter aperture and 0.8\% detection efficiency for $\gamma$ rays, and is capable of recording counts/second in integer units up to a saturation threshold of 100 counts per second.  If the Geiger counter is considered to be ``part of'' $\mathbf{O}$, $\mathbf{O}$ can be regarded as deploying a POVM defined over $\mathcal{H}_{\mathbf{E}}$ that yields as recordable outcome values the integers 1, 2, ... 100, where these outcome values depend, via the response function of the Geiger counter, on the flux of $\gamma$ radiation at $\mathbf{O}$'s location.  Suppose $\mathbf{O}$ is equipped with an effectively-infinite memory and is allowed to wander around the laboratory.  If $\mathbf{O}$ happens to be 1 m from a 100 $\mu$Ci $^{60}$Co source, $\mathbf{O}$ will typically record an outcome of 3.  If $\mathbf{O}$ happens to be 2 m from a 400 $\mu$Ci $^{137}$Cs source, $\mathbf{O}$ will also typically record an outcome of 3.  For $\mathbf{O}$, therefore, a 100 $\mu$Ci $^{60}$Co source located 1 m away is indistinguishable from 400 $\mu$Ci $^{137}$Cs source located 2 m away.  Were $\mathbf{O}$ to be equipped with additional POVMs - for example, ones that reported the direction that the Geiger counter was pointing, $\mathbf{O}$'s own direction of motion, or the distance $\mathbf{O}$ traversed from one recorded observation to the next - $\mathbf{O}$ might be able to distinguish between these two sources of radiation.  If $\mathbf{O}$ is equipped only with the Geiger counter, however, these distinct physical systems can be exchanged arbitrarily without altering $\mathbf{O}$'s observational records.  

As Moore's theorem applies to any finite number of classically-recorded observational outcomes, it guarantees that any observer equipped with only a finite number of POVMs will be in the predicament of an observer equipped only with a Geiger counter: such an observer will never be able to say how many distinct systems her recorded observational outcomes refer to.  In particular, an observer $\mathbf{O}$ that is equipped with a POVM $\lbrace E_{i}^{\mathbf{S}} \rbrace$, defined over all of $\mathbf{O}$'s environment $\mathbf{E}$, that reports states of some collection $\mathbf{S}$ of degrees of freedom within $\mathbf{E}$, but that lacks a POVM that reports the state of some other degree of freedom $\phi$ of $\mathbf{E}$ will be unable to distinguish $\mathbf{S}$ from $\mathbf{S} \otimes \phi$, and hence unable to distinguish interactions with $\mathbf{S}$ from interactions with $\mathbf{S} \otimes \phi$.  If the interaction $H_{\mathbf{S}\phi}$ between $\mathbf{S}$ and $\phi$ is negligible, an exchange of $\mathbf{S} \otimes \phi$ for $\mathbf{S}$ in the environment of $\mathbf{O}$ will have no consequences for future observational outcomes obtained by $\mathbf{O}$ through the use of $\lbrace E_{i}^{\mathbf{S}} \rbrace$.  If the interaction $H_{\mathbf{S}\phi}$ is not negligible, however, an exchange of $\mathbf{S} \otimes \phi$ for $\mathbf{S}$ in $\mathbf{O}$'s environment will potentially have consequences for future observations that employ $\lbrace E_{i}^{\mathbf{S}} \rbrace$.  A fixed observer $\mathbf{O}$ equipped only with a Geiger counter, for example, cannot distinguish a constant, uniform $\gamma$-photon flux $\Phi_{\gamma}$ pervading all of space from a constant point $\gamma$ source at a fixed location if the two sources produce the same event count rate at $\mathbf{O}$'s location.  The latter source, however, has a positional degree of freedom that interacts with event count rate, through which future observational outcomes from $\mathbf{O}$'s location may be affected.  Such an inability to strictly specify the interaction between an observed system and its environment, and hence the inability to strictly specify the POVM implemented by a system employed as an experimental apparatus, can have practical consequences.  The history of science includes many instances in which functionally-significant changes in the degrees of freedom of an experimental apparatus that went undetected by the POVMs employed by observers to identify the apparatus led later to unexpected observational outcomes and reports of surprising phenomena.  Such historical instances indicate that, common assumptions regarding observer-independent ``emergence'' to the contrary \cite{fields:12b}, our universe does not assign any special status or stability to the collections of degrees of freedom identified as ``systems'' by the POVMs that are implemented by human observers.  

Recognition of this observable-dependent exchange symmetry allows Moore's theorem to be restated in a way that appears stronger than the original but is in fact equivalent: no finite sequence of classically-recorded observational outcomes $k_{1} ... k_{N}$ is sufficient to demonstrate that each of the $k_{i}$ was produced by an interaction with the same collection of physical degrees of freedom.  Human observers tend to assume that if something \textit{looks like} the same thing that was observed previously - if it produces the same observational outcome(s) using whatever POVM(s) they have available - then it \textit{is} the same thing that was observed previously.  This application of ``Leibniz's Law'' clearly contributes to the world making sense.  Moore's theorem shows, however, that it is empirically unjustifiable: that two observations are observations of the same system may in many cases be a reasonable theoretical hypothesis, but it can never be regarded as a fact.

\section{Quantum no-go theorems and observable-dependent exchange symmetry}

A steadily-increasing collection of ``no-go'' theorems demonstrate that the acquisition of classical information by observers is severely restricted within quantum theory.  These quantum no-go theorems follow from the assumption of unitary dynamics; they require no special assumptions about the process of observation.  They can each, however, be regarded as blocking particular inferences about the states of quantum systems based on finite observations of those systems.  The inferences blocked by three of the best-known ``no-go'' theorems, the ``no-cloning'' theorem of Wooters and Zurek \cite{wooters:82}, the Kochen-Specker theorem \cite{kochen:67}, and Bell's \cite{bell:64} theorem are considered here.  The no-cloning theorem states that an arbitrary quantum state cannot be precisely duplicated by a unitary operation; hence it blocks the inference that any observed quantum state is a clone of any other quantum state.  As summarized by Mermin \cite{mermin:93} following a comparative review of alternative statements and proofs, the Kochen-Specker theorem states that outcomes obtained by measuring even mutually-commuting observables depend on the manner in which they are measured, while Bell's theorem states that the context-dependence established by the Kochen-Specker theorem characterizes measurements of mutually-commuting observables even when the measurement sites are distant.  The Kochen-Specker theorem thus blocks the inference that the state of a quantum system $\mathbf{S}$ obtained by making measurements with a POVM $\lbrace E_{i}^{\mathbf{S1}} \rbrace$ will be the same as that obtained by making measurements with a POVM $\lbrace E_{i}^{\mathbf{S2}} \rbrace$, even if $\lbrace E_{i}^{\mathbf{S1}} \rbrace$ and $\lbrace E_{i}^{\mathbf{S2}} \rbrace$ are both defined over a single environment $\mathbf{E}$ and report states of some subset $\mathbf{S}$ of the degrees of freedom of $\mathbf{E}$.  Bell's theorem blocks the inference that the outcomes obtained by acting on a spatially-extended collection of degrees of freedom $\mathbf{S}$ with two POVMs $\lbrace E_{i}^{\mathbf{S1}} \rbrace$ and $\lbrace E_{i}^{\mathbf{S2}} \rbrace$ deployed at different locations can be considered to be independent.

Let us consider no-cloning first.  By prohibiting quantum cloning, the no-cloning theorem blocks the assumption that a ``faithful copy'' of a quantum state can be employed to store, confirm, or otherwise access quantum information that is also employed, via the copied original, as an input to some other process.  For such a faithful copy of a quantum state to be of use, it must be possible to observationally access the copy, at some time after its creation, with full confidence that it \textit{is} a copy, even if the original state of which it is a copy has been destroyed.  Moore's theorem clearly blocks the inference that any sequence of classically-recorded observational outcomes obtained from any state $|\psi\rangle$ are sufficient to demonstrate that $|\psi\rangle$ is a faithful copy of any other quantum state, whether the other, ``original'' state has been observed or merely described theoretically.  In particular, observational outcomes produced by any finite collection of POVMs are subject to observable-dependent exchange symmetry, and hence insufficient to distinguish states $|\psi\rangle$ of only the degrees of freedom $\psi$ reported by the deployed POVMs from composite states $|\psi ~ \phi\rangle$ that also involve degrees of freedom $\phi$ undetected by any of the deployed POVMs.  As discussed above, such a composite state $|\psi ~ \phi\rangle$ may display behavior that is unpredictable from knowledge of $|\psi\rangle$ alone.  Hence Moore's theorem prevents any quantum state, regardless of its provenance, from being regarded on the basis of finite observations as a clone of any other quantum state.  The strongest statement permitted by Moore's theorem is that two quantum states are indistinguishable by a finite set of observations that have either been performed or been simulated theoretically; this weaker statement is clearly insufficient for ``quantum cloning.''  Hence while Moore's theorem does not prohibit quantum cloning dynamically - Moore's theorem does not concern dynamics at all - it prohibits treating any quantum state as a clone; hence the consequences Moore's theorem for any operations involving quantum states include the consequences of the no-cloning theorem as a special case.

The Kochen-Specker theorem blocks the inference that two POVMs $\lbrace E_{i}^{\mathbf{S1}} \rbrace$ and $\lbrace E_{i}^{\mathbf{S2}} \rbrace$ will produce the same sequence of classically-recorded observational outcomes $k_{1} ... k_{N}$ even if they act on the same degrees of freedom within a single subset of degrees of freedom $\mathbf{S}$.  Let us ask: under what circumstances \textit{would} this inference be justified?  Two POVMs $\lbrace E_{i}^{\mathbf{S1}} \rbrace$ and $\lbrace E_{i}^{\mathbf{S2}} \rbrace$ can be expected to produce the same sequence of classically-recorded observational outcomes $k_{1} ... k_{N}$ only if it is known that they both act \textit{only} on a particular set of degrees of freedom $\lbrace \psi, \phi, ... \chi \rbrace$, i.e. only on the Hilbert space spanned by $\lbrace \psi, \phi, ... \chi \rbrace$.  Observable-dependent exchange symmetry prevents this from ever being known about any physically-implemented POVM; no finite sequence of observations can determine, for any POVM, the complete set of degrees of freedom for which that POVM yields finite outcomes.  Hence for physically-implemented POVMs - the only POVMs of relevance to observers - Moore's theorem already prohibits observers from making the inference that is prohibited by the Kochen-Specker theorem.  As in the case of the no-cloning theorem, the Kochen-Specker prohibition is based on a dynamical assumption - in particular, the presence of entanglement - while Moore's prohibition is based on a consideration of what can be inferred from measurements even in a classical setting.

Bell's theorem blocks the assumption that two POVMs $\lbrace E_{i}^{\mathbf{S1}} \rbrace$ and $\lbrace E_{i}^{\mathbf{S2}} \rbrace$ act independently on spatially-distant components of a single system $\mathbf{S}$.  Let us again ask for the circumstances under which such an assumption would be justified.  Two POVMs $\lbrace E_{i}^{\mathbf{S1}} \rbrace$ and $\lbrace E_{i}^{\mathbf{S2}} \rbrace$ can be expected to produce independent sequences of outcomes $k_{1} ... k_{N}$ and $l_{1} ... l_{N}$ only if they act on sets of degrees of freedom $\lbrace \psi, \phi, ... \chi \rbrace$ and $\lbrace \psi^{\prime}, \phi^{\prime}, ... \chi^{\prime} \rbrace$ that are separable within the Hilbert space $\mathcal{H}_{\mathbf{E}}$ over which $\lbrace E_{i}^{\mathbf{S1}} \rbrace$ and $\lbrace E_{i}^{\mathbf{S2}} \rbrace$ are defined.  Observable-dependent exchange symmetry prevents this from being known on the basis of observations about any pair of physically-implemented POVMs.  The presumed spatial separation between the locations at which the operators are deployed is, moreover, irrelevant to this conclusion.  The two sets of outcomes $k_{1} ... k_{N}$ and $l_{1} ... l_{N}$ must be jointly recorded at some location in order for the question of their independence to arise; if the outcomes are obtained at different locations by different observers as in the usual Bell's theorem scenario, one or both of the observers must communicate the outcomes obtained either to the other or to some third party who records them both.  If information must be physically encoded at all times, a physical process must deliver the communicated outcome(s) to the recording location.  In this case the system $\mathbf{S}$ can be notionally expanded within $\mathbf{E}$ to include this physical information-delivery process, and the operators $\lbrace E_{i}^{\mathbf{S1}} \rbrace$ and $\lbrace E_{i}^{\mathbf{S2}} \rbrace$ can be regarded as acting on this expanded system at the location where the outcomes are jointly recorded.  

Recognizing the restrictions that are imposed on inferences from classically-recorded observational outcomes by the no-cloning theorem, the Kochen-Specker theorem and Bell's theorem as restrictions already imposed on such inferences by Moore's theorem leads naturally to the question of why this should be so.  The no-cloning theorem, the Kochen-Specker theorem and Bell's theorem all follow from the assumption of unitary dynamics and hence of the possibility - indeed inevitability - of entanglement between interacting systems.  These theorems have no analogs in classical mechanics, in which interacting systems are assumed to be separable at all times.  Moore's theorem, on the other hand, makes no explicit assumptions about the dynamics of the systems being observed; it only assumes only that the space of all possible classical FSMs can be considered to be well-defined.  Hence Moore's theorem does not contradict classical mechanics by implying that classical systems are non-separable.  It only restricts the ability of observers to \textit{identify} classical systems, and in particular, to place upper bounds on their degrees of freedom.  Hence what it contradicts is an implicit assumption of classical physics that systems are ``transparent'' to observation, that they have only the degrees of freedom that they are observed to have. 

How Moore's theorem subjects even classical observations to the prohibitions associated with the quantum no-go theorems can be made clear by translating the assumption that the space of all possible classical FSMs can be considered to be well-defined from mathematical to physical terms.  If Moore's ``machines'' are regarded as physically implemented, then Moore's inference that ``there exist other machines experimentally distinguishable from $\mathbf{S}$'' requires the actual existence of other physical systems that have physical dynamics that duplicate those of $\mathbf{S}$ to whatever extent the dynamics of $\mathbf{S}$ have been observed, but that can diverge from those of $\mathbf{S}$ with the very next observation.  With this physical interpretation, Moore's theorem becomes the statement that any finitely-recorded observational history is consistent with arbitrarily many futures that reveal the dynamical effects of degrees of freedom that were always present, but were previously ``hidden'' in the straightforward sense of being not yet detected.  The notion that a fixed observational history is consistent with arbitrarily many distinct futures is a familiar one: it is the central notion of Everett's \cite{everett:57} model of the universe as a single Hilbert space in which unitary dynamics unfolds.  From the perspective of Moore's theorem, ``systems'' are defined solely by observational histories, and hence correspond to ``branches'' in a fully-unitary, Everettian conception of quantum theory.  Hence it is unsurprising that conclusions reached from Moore's theorem and from the assumption of unitary dynamics should be similar.

\section{Conclusion}

In a passage typical of many others, Schlosshauer characterizes classical physics as follows:
\begin{quote}
``Here (i.e. in classical physics) we can enlarge our `catalog' of physical properties of the system (and therefore specify its state more completely) by performing an arbitrary number of measurements of identical physical quantities, in any given order.  Moreover, many independent observers may carry out such measurements (and agree on the results) without running into any risk of disturbing the state of the system, even though they may have been initially completely ignorant of this state.''
\begin{flushright}
(\cite{schloss:07} p. 16)
\end{flushright} 
\end{quote}
To be ``completely ignorant'' of the state of a system is to not know the values of any of the system's state variables: to not know its size, shape, location, direction of travel, or anything else about it.  How, then, are observers to \textit{identify} the system as a target of observations, much less agree that they have observed the \textit{same} system?  To be identifiable as a target of observations by ``completely ignorant'' observers, a system must be taken as \textit{given}: given by the laws of physics, the initial conditions of the universe, or some omnipotent power.  As characterized by Schlosshauer, classical physics takes the systems composing the world as given.  This assumption is familiar from \S 2: it is the assumption that any particular set of observational outcomes can be attributed, by definition, to a particular collection of physical degrees of freedom.

If dynamics are unitary and universal, one cannot take systems as given.  One can choose to ignore degrees of freedom that are not explicitly represented by one's classically-recorded observational outcomes, and one can trace out such degrees of freedom when performing decoherence calculations, but one must remain aware that doing so has no effect on their causal relevance in reality.  Both Moore's theorem and the quantum no-go theorems are continual reminders of this fact.

\section*{Acknowledgment}

Thanks to Louis Vervoort and to an anonymous referee for close readings and helpful comments.

\end{document}